\documentclass[proof]{WileyASNA-v1}

\articletype{Article Type}%

\received{26 April 2016}
\revised{6 June 2016}
\accepted{6 June 2016}

\usepackage{textcomp}

\begin{document}

\title{Evaluation of scientific CMOS sensors for sky survey applications}

\author[1]{S. Karpov*}

\author[1]{A. Bajat}

\author[1]{A. Christov}

\author[1]{M. Prouza}

\authormark{S.~Karpov \textsc{et al}}

\address[1]{\orgdiv{CEICO}, \orgname{Institute of Physics, Czech Academy of Sciences}, \orgaddress{\state{Prague}, \country{Czech Republic}}}

\corres{*Sergey Karpov, Institute of Physics of Czech Academy of Sciences, Na Slovance 1999/2, 182 21 Praha 8, Czech Republic \email{karpov.sv@gmail.com}}

\presentaddress{Sergey Karpov, Institute of Physics of Czech Academy of Sciences, Na Slovance 1999/2, 182 21 Praha 8, Czech Republic}

\abstract{
  Scientific CMOS image sensors are a modern alternative for a typical CCD detectors, as they offer both low read-out noise, large sensitive area, and high frame rates. All these makes them promising devices for a modern wide-field sky surveys. However, the peculiarities of CMOS technology have to be properly taken into account when analyzing the data. In order to characterize these, we performed an extensive laboratory testing of Andor Marana sCMOS camera. Here we report its results, especially on the temporal stability and linearity, and compare it to the previous versions of Andor sCMOS cameras. We also present the results of an on-sky testing of this sensor connected to a wide-field lens, and discuss its applications for an astronomical sky surveys.
  }

\jnlcitation{\cname{%
\author{S. Karpov},
\author{A. Bajat},
\author{A. Christov},
\author{M. Prouza}} (\cyear{2019}),
\ctitle{Scientific CMOS sensors for sky surveys}, \cvol{2019;00:1--6}.}

%% \fundingInfo{Funding info text.}

\maketitle

% \footnotetext{\textbf{Abbreviations:} ANA, anti-nuclear antibodies; APC, antigen-presenting cells; IRF, interferon regulatory factor}

\section{Introduction}\label{sec_intro}

Sky survey applications require large format image sensors with high quantum efficiency, low read-out noise, fast read-out and a good inter- and cross-pixel stability and linearity. Charge-Coupled Devices (CCDs), typically employed for such tasks, lack only the read-out speeds, which significantly lowers their performance for detecting and characterizing rapidly varying or moving objects. On the other hand, Complementary Metal–Oxide–Semiconductor (CMOS) imaging sensors, widely used on the consumer market, typically displays levels of read-out noise unacceptably large for precise astronomical tasks (tens of electrons), as well as significantly worse uniformity and stability than CCDs.

However, recent development in the low-noise CMOS architectures (see e.g. \citet{spie_scmos}) allowed to design and create a market-ready large-format (2560x2160 6.5$\mu$m pixels) CMOS chips with read-out noise as low as 1-2 electrons, on par with best CCDs \citep{spie_neo} -- so-called ``scientific CMOS'' (sCMOS) chips. Like standard CMOS sensors (and unlike CCDs), they did not perform any charge transfer between adjacent pixels, employing instead individual column-level amplifiers with parallel read-out and dual 11-bit analog-to-digital converters (ADCs) operating in low-gain and high-gain mode, correspondingly, and an on-board field-programmable gate array (FPGA) logic scheme that reconstructs a traditional 16-bit reading for every pixel from two 11-bit ones.

The cameras based on this original sCMOS chip, CIS2051 by Fairchild Imaging, have been available since 2009. Andor Neo is one of such cameras and is currently widely used in astronomical applications, especially for tasks that require high frame rates like satellite tracking \citep{schildknecht_2013}, fast photometry \citep{qui_2013} or rapid optical transients detection \citep{karpov_2019}. Detailed characterization of this camera presented in these works displays its high performance and overall good quality of delivered data products, with the outlined problems related mostly to the non-linearity at and above the amplifier transition region around 1500 ADU and a low full well depth of about only 20k electrons.

Scientific CMOS sensors with larger well depth, up to 120k electrons \citep{spie_400}, and a back-illuminated options having significantly better quantum efficiency (up to 95\%) have been released later by a GPixel. Andor Marana \citep{andor_marana} is a camera built around such back-illuminated chip, GSense400BSI. We decided to perform the laboratory and on-sky characterization of this recently released camera, generously provided to us by manufacturer for testing, in order to compare its properties with the ones of earlier scientific CMOSes (see Table~\ref{tab_marana} for a brief comparison of parameters of Andor Marana to Andor Neo used here as a representative of a first-generation sCMOS) and to assess its performance for sky survey applications. Such a device, if proven to be stable enough, may be of extreme importance for a tasks of precise photometry in wide field sky surveys, especially when high temporal resolution is desirable, e.g. for the detection and study of rapid optical transients \citep{karpov_2010, karpov_2019}, space debris tracking \citep{karpov_2016} or observations of faint meteors \citep{karpov_meteors_2019}. Due to large frame format, absence of microlens raster on top of the chip, good quantum efficiency and fast read-out, such device may also be a promising detector for a next generation of FRAM atmospheric monitoring telescopes \citep{fram,fram_cta}.

% The comparison of recent Andor Marana sCMOS to previous model, Andor Neo, according to manufacturer specifications is shown in Table~\ref{tab_marana}. It is evident that it has significantly larger dark current

\begin{center}
\begin{table}[t]%
\centering
\caption{Comparison of parameters of Andor Marana sCMOS camera, studied in this work, to the ones of Andor Neo camera.
\label{tab_marana}}%
\tabcolsep=0pt%
\begin{tabular*}{20pc}{@{\extracolsep\fill}lcc@{\extracolsep\fill}}
\toprule
\textbf{} & \textbf{Andor Neo}  & \textbf{Andor Marana} \\
  \midrule
  Chip & Front illum. & Back illum. \\
  Format & 2560 x 2160 & 2048 x 2048 \\
  Diagonal & 21.8 mm & 31.8 mm \\
  Pixel size  & 6.4 $\mu$m & 11 $\mu$m \\
  Peak QE & 60\% & 95\% \\
  Shutter & Rolling, Global & Rolling \\
  Max FPS & 30 & 48 \\
  Read noise & 1.0 e$^-$ & 1.6 e$^-$ \\
  Dark current,  e$^-$/pix/s & 0.007@-40$^\circ$C & 0.2@-45$^\circ$C \\
  Full well depth  & 30000 e$^-$ & 85000 e$^-$ \\
  Interface & CameraLink & USB3.0 \\
\bottomrule
\end{tabular*}
% \begin{tablenotes}
% \item Source: Example for table source text.
% \item[$\dagger$] Example for a first table footnote.
% \item[$\ddagger$] Example for a second table footnote.
% \end{tablenotes}
\end{table}
\end{center}

\section{Experimental setup}

\begin{figure}[t]
  \centerline{\resizebox*{1.0\columnwidth}{!}{\includegraphics[angle=0]{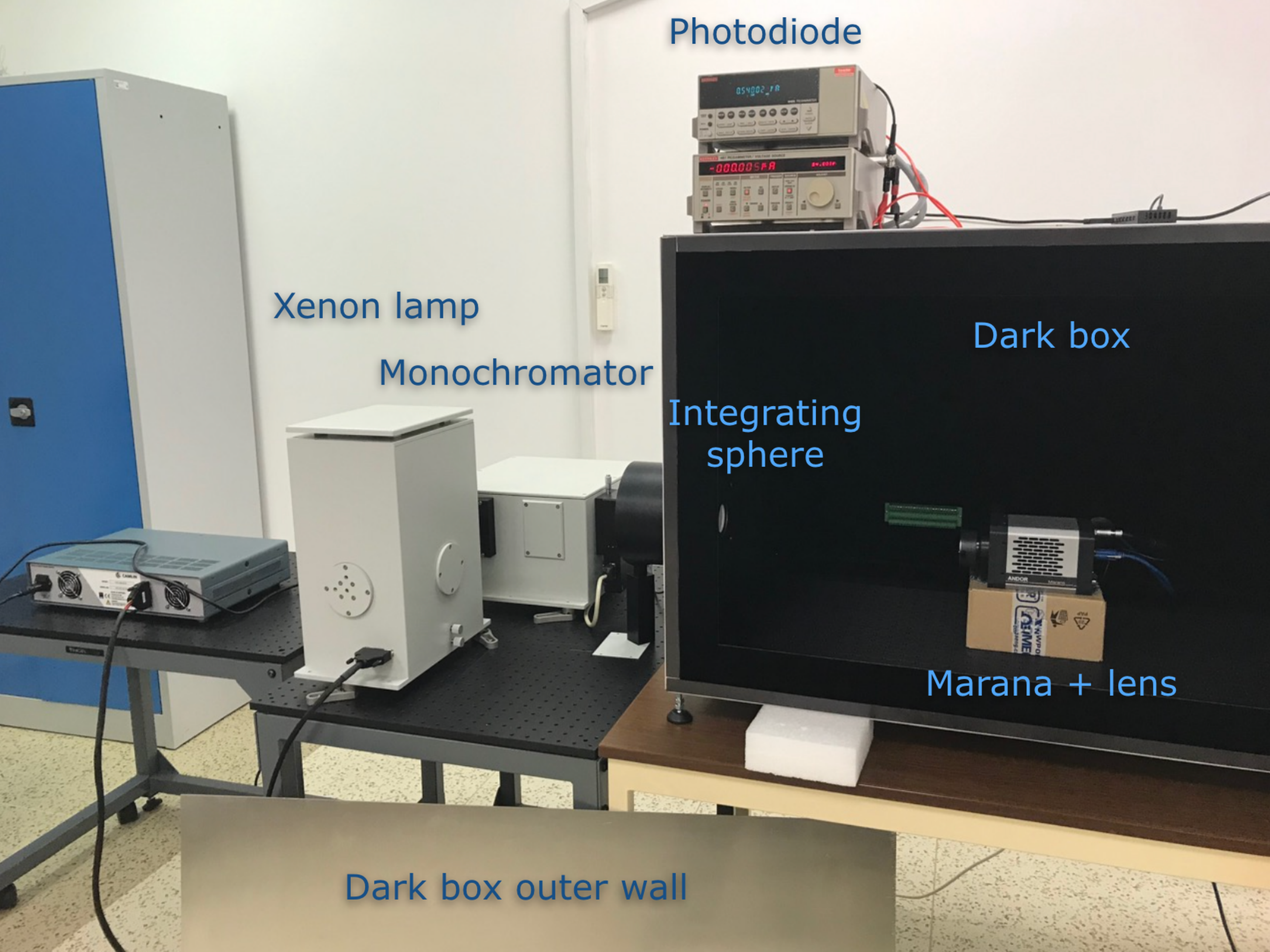}}}
  \caption{Laboratory setup used for testing Andor Marana sCMOS camera at Institute of Physics, Czech Academy of Sciences in Mar 2019. The side wall of a dark box is open so the camera inside it is visible. The photodiode used to control light source intensity is installed in a side port of the integrating sphere, and is measured by a dedicated picoammeter.
    \label{fig_lab}}
\end{figure}

\begin{figure*}[t]
  \centerline{
    \resizebox*{0.66\columnwidth}{!}{\includegraphics[angle=0]{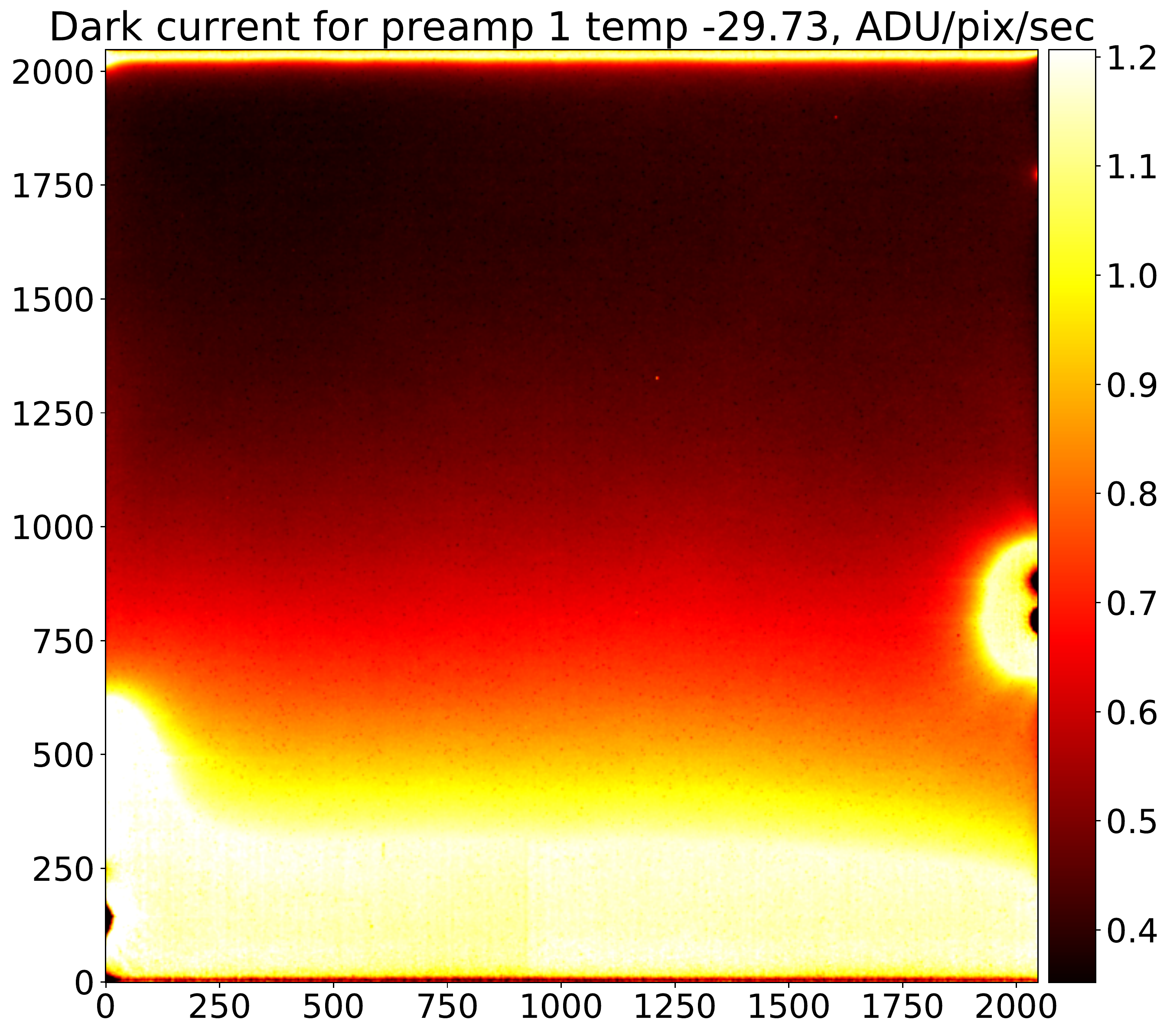}}
    \resizebox*{0.65\columnwidth}{!}{\includegraphics[angle=0]{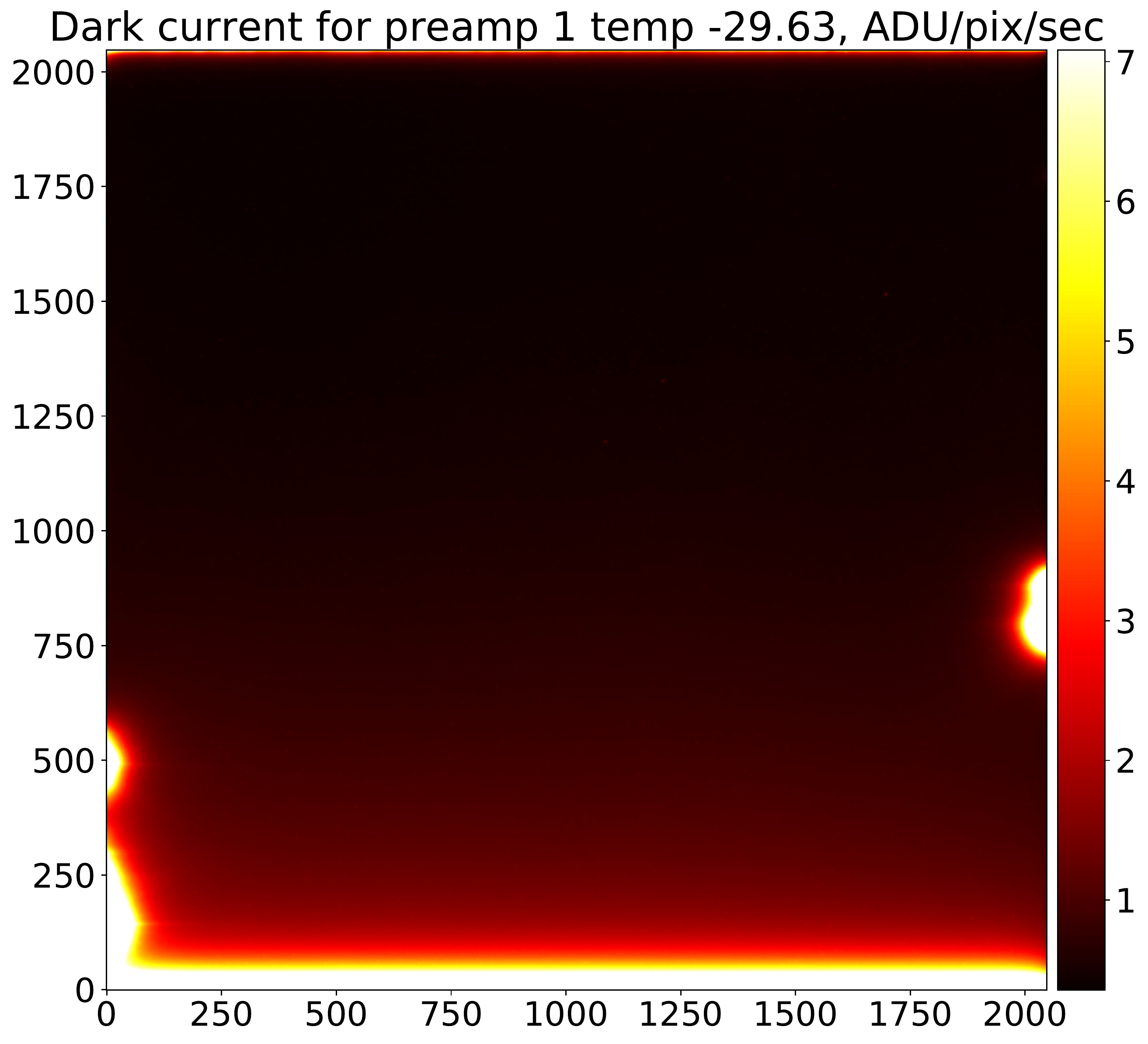}}
    \resizebox*{0.675\columnwidth}{!}{\includegraphics[angle=0]{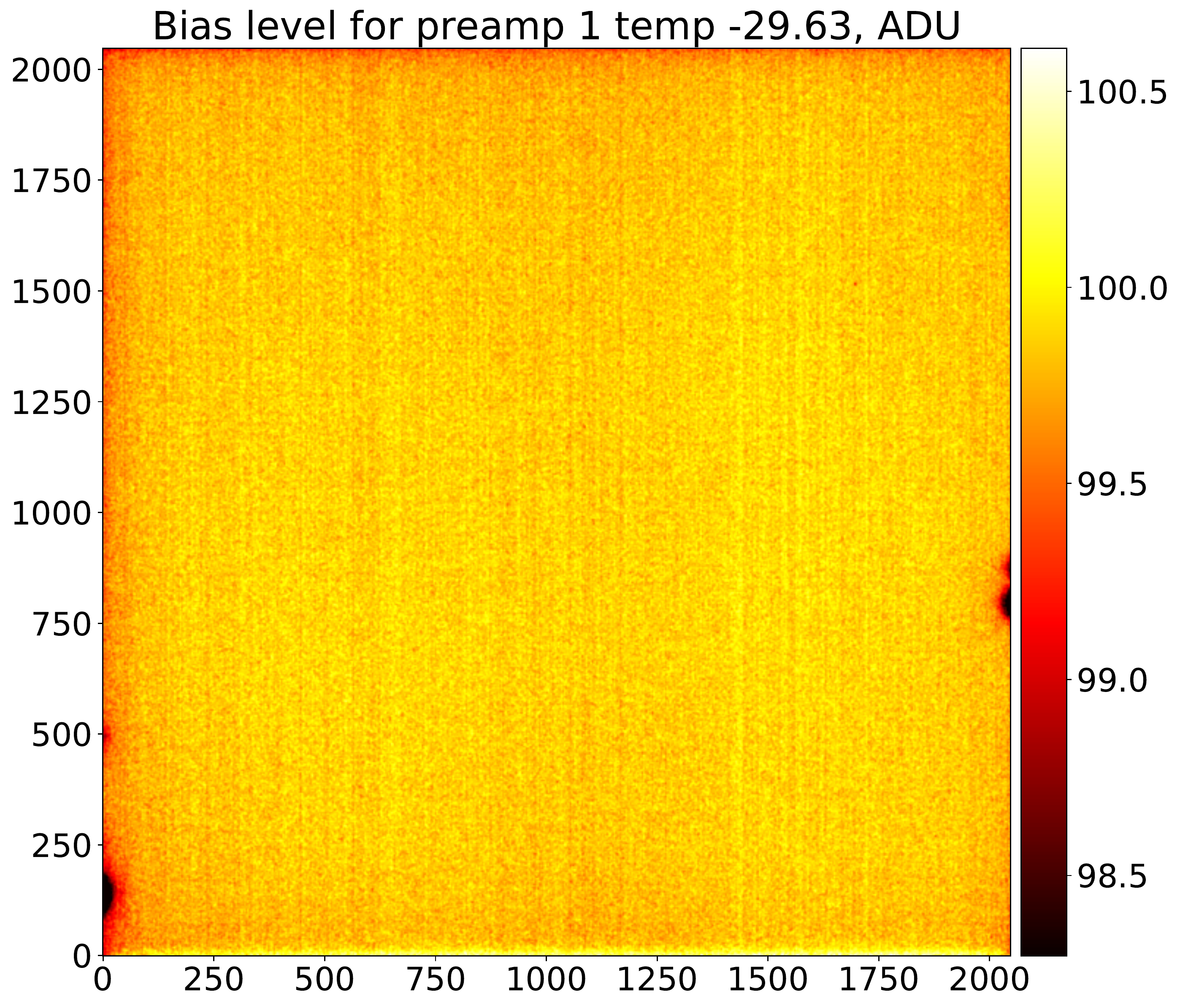}}
  }
  \caption{Maps of a dark current for a default camera regime with Anti-Glow correction (left, median value is 0.55 ADU/pixel/s) and with correction disabled (middle, median value is 0.84 ADU/pixel/s), as well as of bias level (right).
    \label{fig_darks}}
\end{figure*}

For testing the camera, we used an existing experimental equipment available at a laboratory of characterization of optical sensors for astronomical applications at Institute of Physics of Czech Academy of Sciences (see Figure~\ref{fig_lab}). That included a fully light-isolating dark box, CAMLIN ATLAS 300 monochromator, CAMLIN APOLLO X-600 Xenon lamp, and an integrating sphere mounted directly on the input port of a dark box. A dedicated photodiode coupled with Keithley picoammeter is used to control the intensity of light inside the integrating sphere. The whole system is controlled by a dedicated {\sc CCDLab} software \citep{ccdlab}, which performs real-time monitoring of a system state and stores it to a database for analyzing its evolution, displays it in a user-friendly web interface, and also allows easy scripting control of its operation. The camera itself was controlled by a {\sc FAST} data acquisition software \citep{fast} specifically designed for operating fast frame rate scientific cameras of various types.

As the laboratory environment we used was not dust-free, the camera was equipped with a Nikkor 300 f/2.8 lens, adjusted in such way as to provide illumination of the whole chip with the light from integrating sphere output window. Due to lens vignetting, this resulted in a slightly bell-shaped flat fields. The same lens has been later used for the on-sky testing of the camera photometric performance. For it, the camera with lens were installed on a Software Bisque' Paramount ME mount, controlled with {\sc RTS2} software \citep{rts2}. The field of view of the camera in this setup was 4.26$^{\circ}$x4.26$^{\circ}$ with 7.5$''$/pixel scale. The whole setup was installed in the dome of FRAM telescope being tested and commissioned at the same time at the backyard of Institute of Physics of Czech Academy of Sciences in Prague.

\section{Laboratory testing}\label{sec_laboratory}

The laboratory testing consisted by a scripted set of imaging sequences of various exposures and durations, acquired under different light intensities or in the dark, and with different readout settings. Most of the sequences were acquired with chip temperature set to -30$^\circ$C, which the camera' Peltier cooler was able to continuously support within the closed area of the dark box with just an air cooling during the whole duration of our experiments. However, for some tests the temperature was also varied. The majority of tests have been performed with a high gain (16-bit) regime, as a most convenient for astronomical applications.

Over every sequence, pixel-level mean values (``mean frames'') and standard deviations (``standard deviation frames'') have been constructed; over some of the sequences, pixel values on individual frames have also been specifically studied.

\subsection{Dark current}\label{sec_dark}

\begin{figure}[t]
  \centerline{\resizebox*{1.0\columnwidth}{!}{\includegraphics[angle=0]{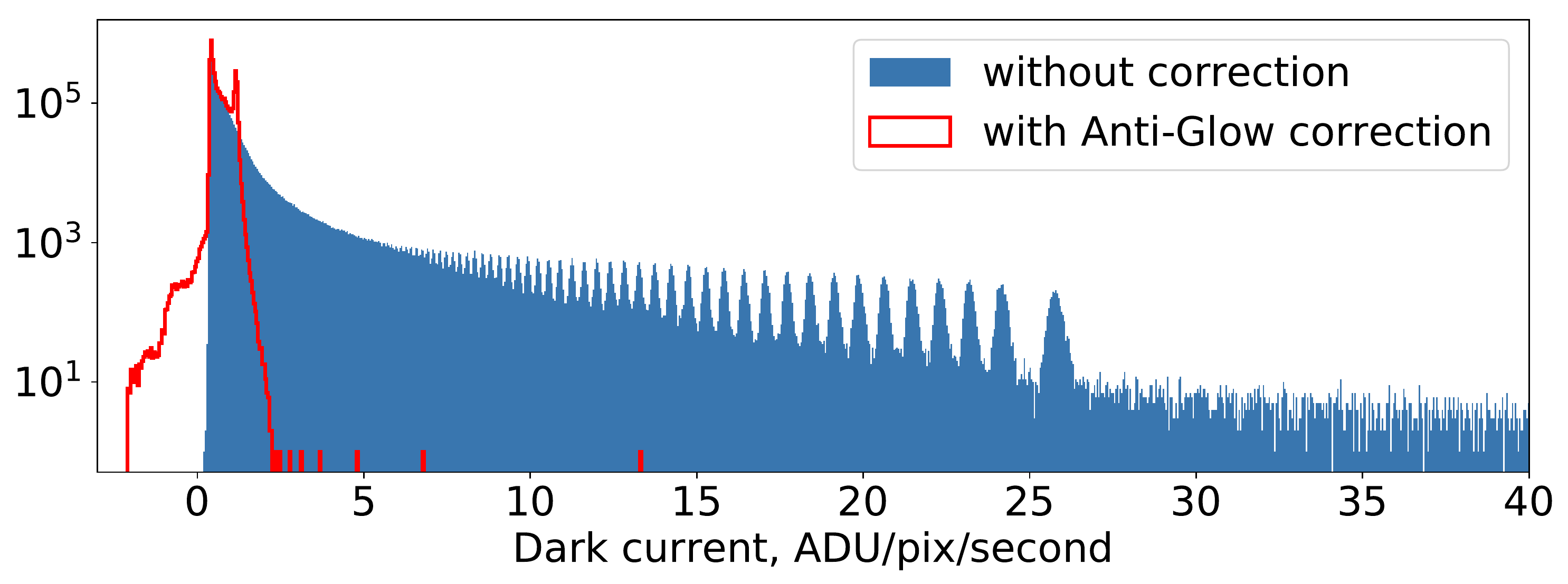}}}
  \caption{Histograms of a per-pixel dark current in default camera regime (with Anti-Glow correction) and with correction disabled.
    \label{fig_dark_hist}}
\end{figure}

We studied the dark current of the camera by acquiring a series of ``dark'' (i.e. with the shutter on monochromator
closed to block the light going into integrating sphere) frames with varying exposures. Then the mean value of every pixel was regressed versus exposure time in order to determine both bias level and dark current on a per-pixel basis. The maps of these values are shown in Figure~\ref{fig_darks}. The dark current shows quite significant edge glow towards both top and bottom edges of the sensor, as well as an extended hot spots along the vertical edges. Interestingly, for 6429 pixels (0.15\% of all) formally measured dark current is negative, as the dark pixel value linearly drops with increased exposure until reaches some fixed level. We attribute it to the occasional over-compensation of a dark current due to  application of ``Anti-Glow technology'' algorithms during on-board frame processing. Turning off this correction (using an undocumented Software Development Kit (SDK) option generously provided to us by an Andor representative) removed such negative values from the dark current map and significantly increased the amplitude of edge glow spots, leading to a long tail in the histogram of its values (see Figure~\ref{fig_dark_hist} for a comparison of histograms with and without glow correction) and slightly increasing the dark current median value (from 0.55 ADU/pixel/s to 0.84 ADU/pixel/s).

The linear slope of over-compensated pixels suggest that the Anti-Glow correction is a linear subtraction of a some pre-computed map multiplied by exposure time, and is therefore equivalent to standard astronomical procedure of dark current correction. Therefore, we suggest disabling this correction when using the camera in a properly calibrated environment in order to avoid possible problems due to over-compensation.

\subsection{Photon transfer curve and linearity}\label{sec_pts}

\begin{figure*}[t]
  \centerline{
    \resizebox*{1.0\columnwidth}{!}{\includegraphics[angle=0]{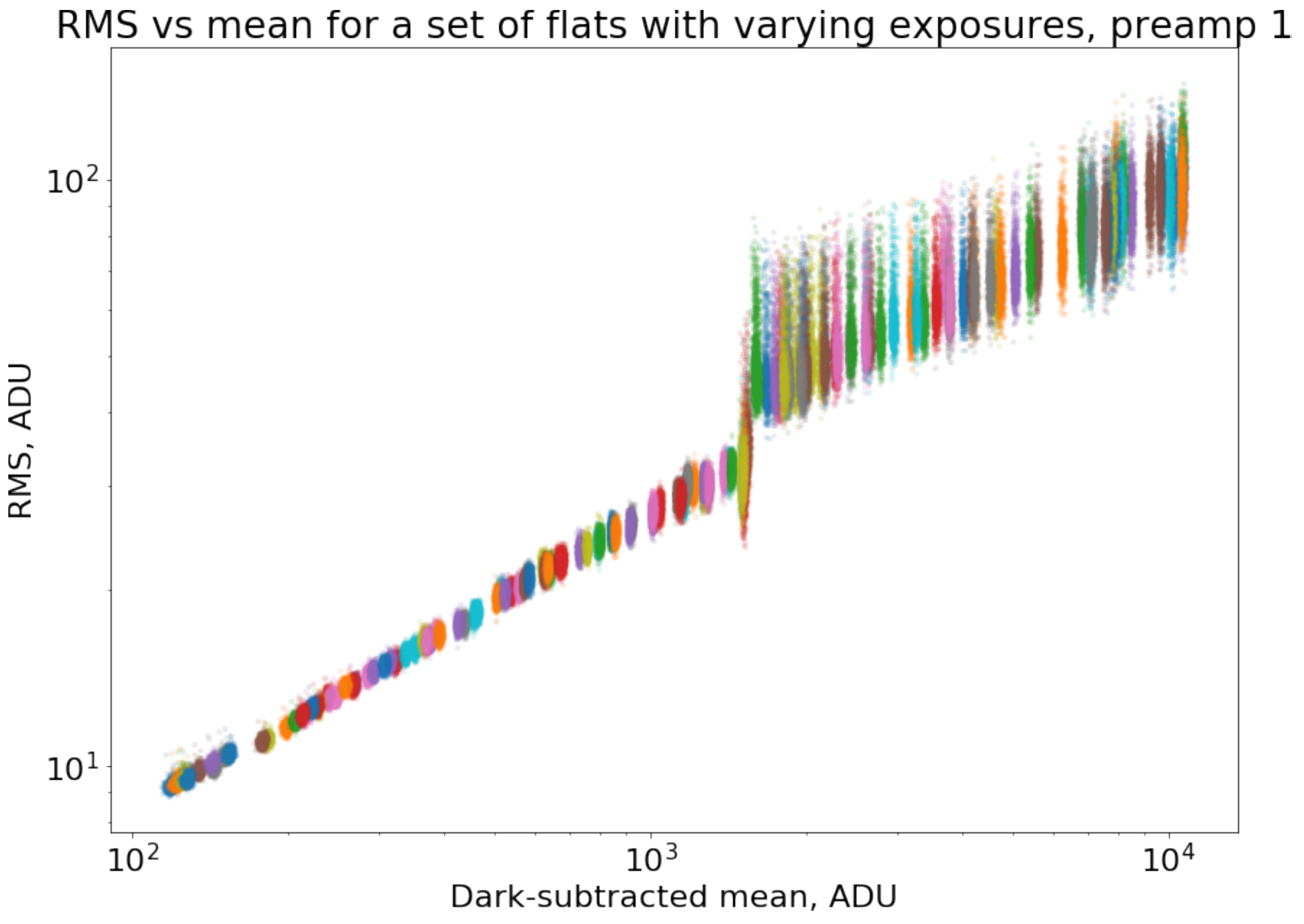}}
    \resizebox*{1.0\columnwidth}{!}{\includegraphics[angle=0]{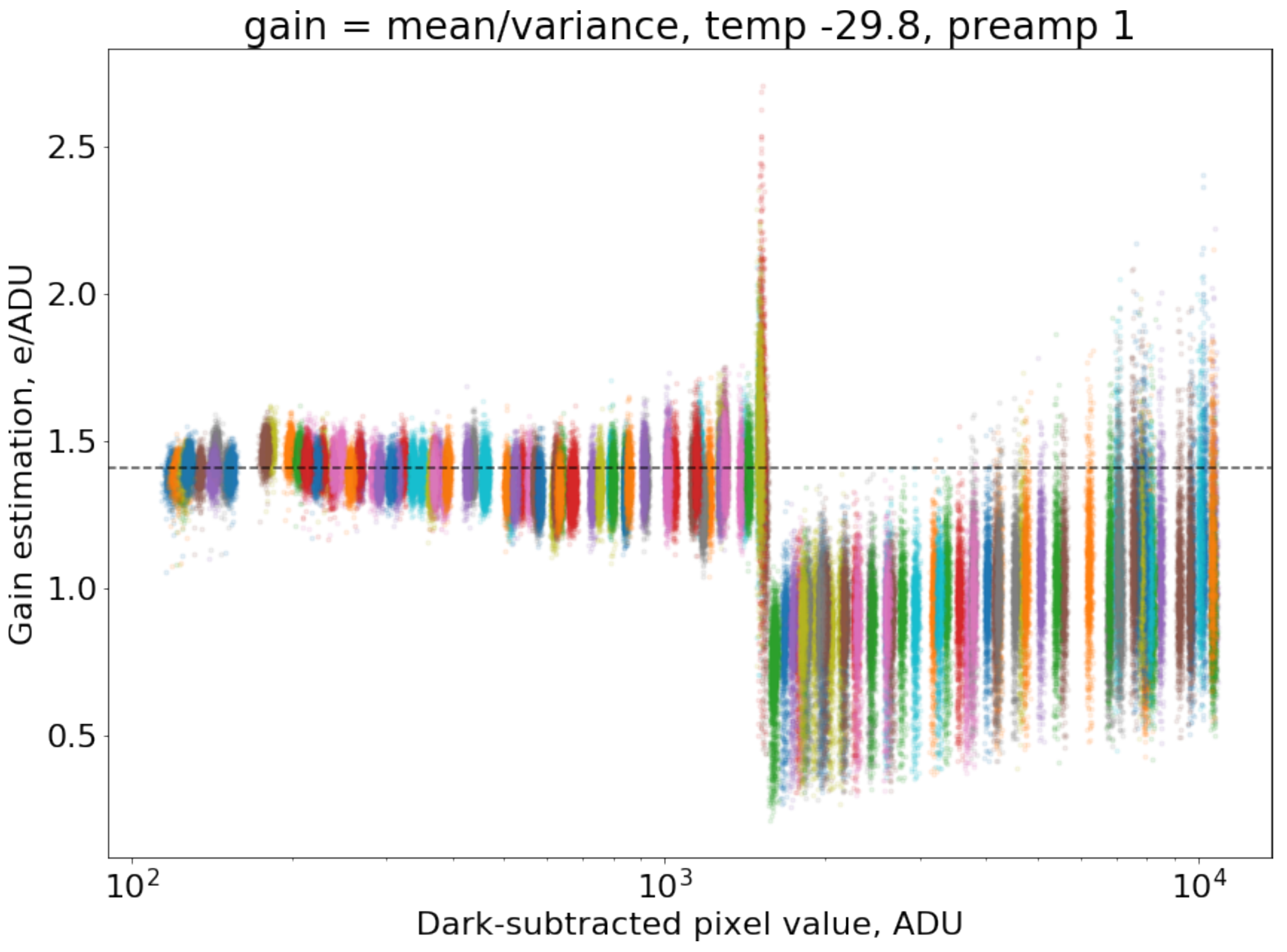}}
  }
  \caption{Left panel -- photon transfer curve, i.e. dependence of a pixel temporal variance on (dark level subtracted) mean value. Right panel -- gain estimated from photon transfer curve as a function of pixel value. On both panels, different colors represent different sequences of frames with varying exposures, and the spread of points of same color -- the difference of corresponding values across different pixels of the sensor. Dashed horizontal line on right panel represents the manufacturer-provided gain level of 1.41 e$^-$/ADU. The jump at around 1500 ADU represents the transition between low-gain and high-gain amplifiers, providing different effective gain and having different spatial structures.
    \label{fig_ptc}}
\end{figure*}

\begin{figure}[t]
  \centerline{\resizebox*{1.0\columnwidth}{!}{\includegraphics[angle=0]{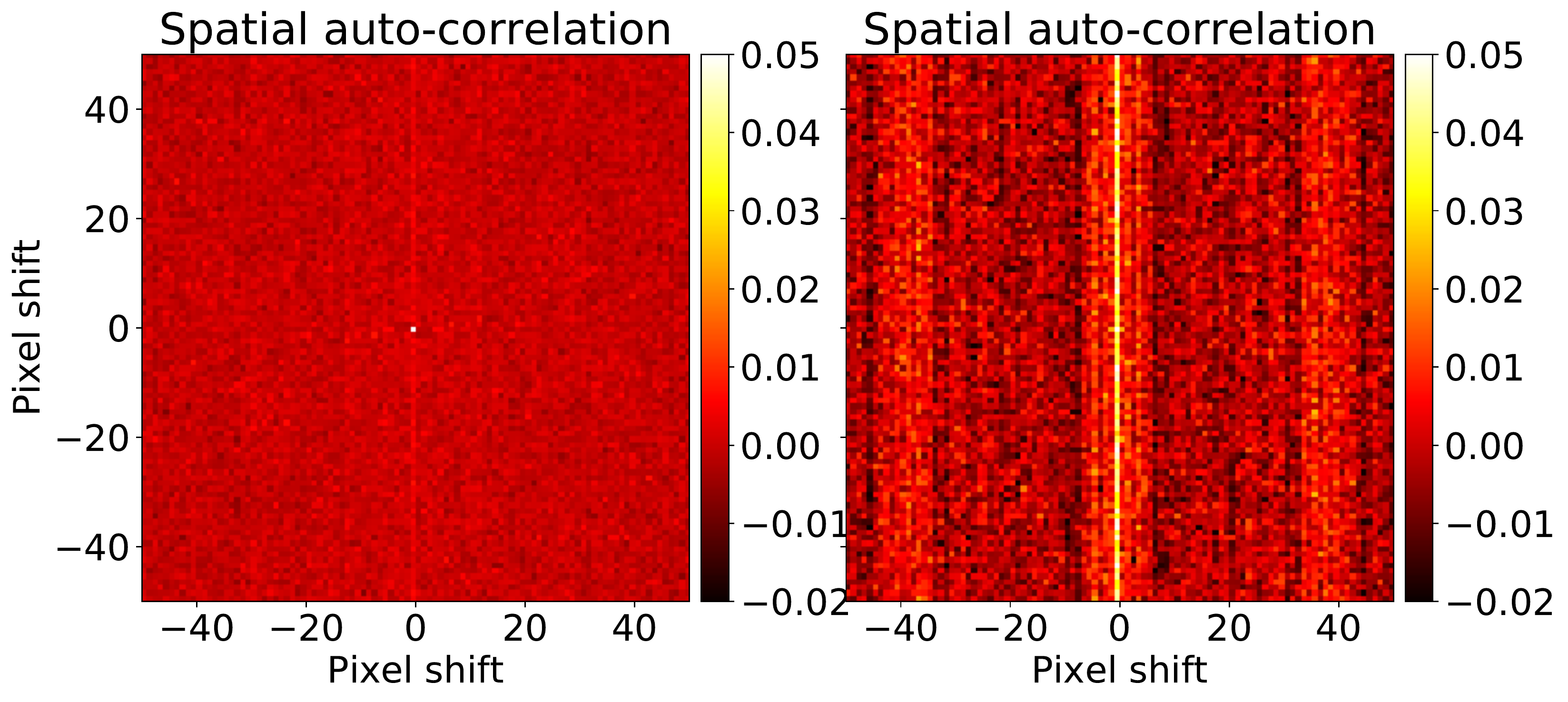}}}
  \caption{Mean spatial autocorrelation for the gain maps computed for mean values between 500 and 1000 ADU (left panel) and between 2500 and 5000 ADU (right panel). The latter shows a distinct column-like structure, reflecting different effective gains for different column level amplifiers.
    \label{fig_gain_corr}}
\end{figure}

\begin{figure}[t]
  \centerline{\resizebox*{1.0\columnwidth}{!}{\includegraphics[angle=0]{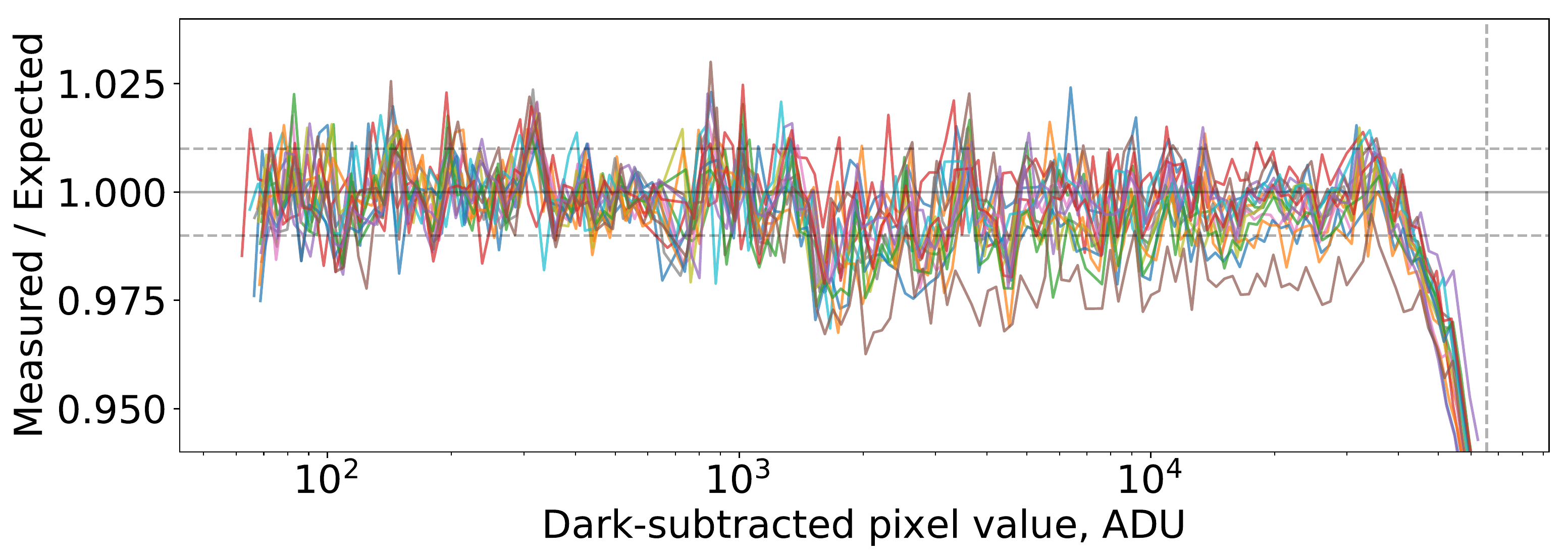}}}
  \caption{Linearity curve for a random set of pixels. The curve represent the ratio of an actually measured signal to the one expected for a linear signal scaling with exposure time, with inverval below 1000 ADU used to define a linear slope. Dashed vertical line to the right marks the position of a digital saturation (65535 ADU), systematic significant deviations from linearity start at about half of this value. The amplifier transition region is easily visible at around 1500 ADU, with the jump amplitude of typically less than couple of percents.
    \label{fig_linearity}}
\end{figure}

In order to build the photon transfer curve (PTC, the dependence of pixel RMS value on its mean, see \citet{ptbook}), we acquired a series of frames with varying exposure time during constant illumination. Due to slightly different properties of individual pixel circuits, column level amplifiers and ADCs, we expect the properties to vary on a pixel to pixel basis, therefore we did not perform any spatial averaging, using instead just a temporal mean and variance of every pixel readings. The resulting PTC is shown in Figure~\ref{fig_ptc}, along with the effective gain computed from it. The sharp jump is seen around 1500 ADU in both plots there, corresponding to the transition between low-gain and high-gain amplifiers. The gain below the transition nicely corresponds to the one reported by manufacturer; above the transition, the effective gain drops by nearly two times. Also, the spatial scatter of gain values across the sensor changes -- from nearly uniform below the transition to a significantly varying ones above it. Spatial auto-correlation shown in Figure~\ref{fig_gain_corr} confirms it, also suggesting that the scatter actually reflects a bit different gain settings for different column level amplifiers.

Linearity curve shown in Figure~\ref{fig_linearity}, on the other hand, does not show any significant jump of comparable amplitude at the transition region, with the slight discontinuity there on the level typically below 2\%, which has a character of a small multiplicative coefficient. Above half of dynamic range, however, there is a systematic change of linearity slope, which may reach up to 10\% towards the saturation point. Overall, the linearity of Marana camera seems a bit better than the one seen in older Andor Neo, where the slope of linearity curve changed above the transition region, with the non-linearity reaching up to 5\% \citep{karpov_2019}.

This all suggests that the properties of flat fields might be a bit different when acquired at intensities below and above the amplifier transition (i.e. 1500 ADU), and requires a more detailed study if one is aimed at precise photometric applications at different intensity levels.

\subsection{Pixel noise and stability}\label{sec_pixelstability}

\begin{figure}[t]
  \centerline{
    % \resizebox*{0.5\columnwidth}{!}{\includegraphics[angle=0]{fig_cmos_neo_corr}}
    % \resizebox*{0.5\columnwidth}{!}{\includegraphics[angle=0]{fig_cmos_marana_corr}}
    \resizebox*{1.0\columnwidth}{!}{\includegraphics[angle=0]{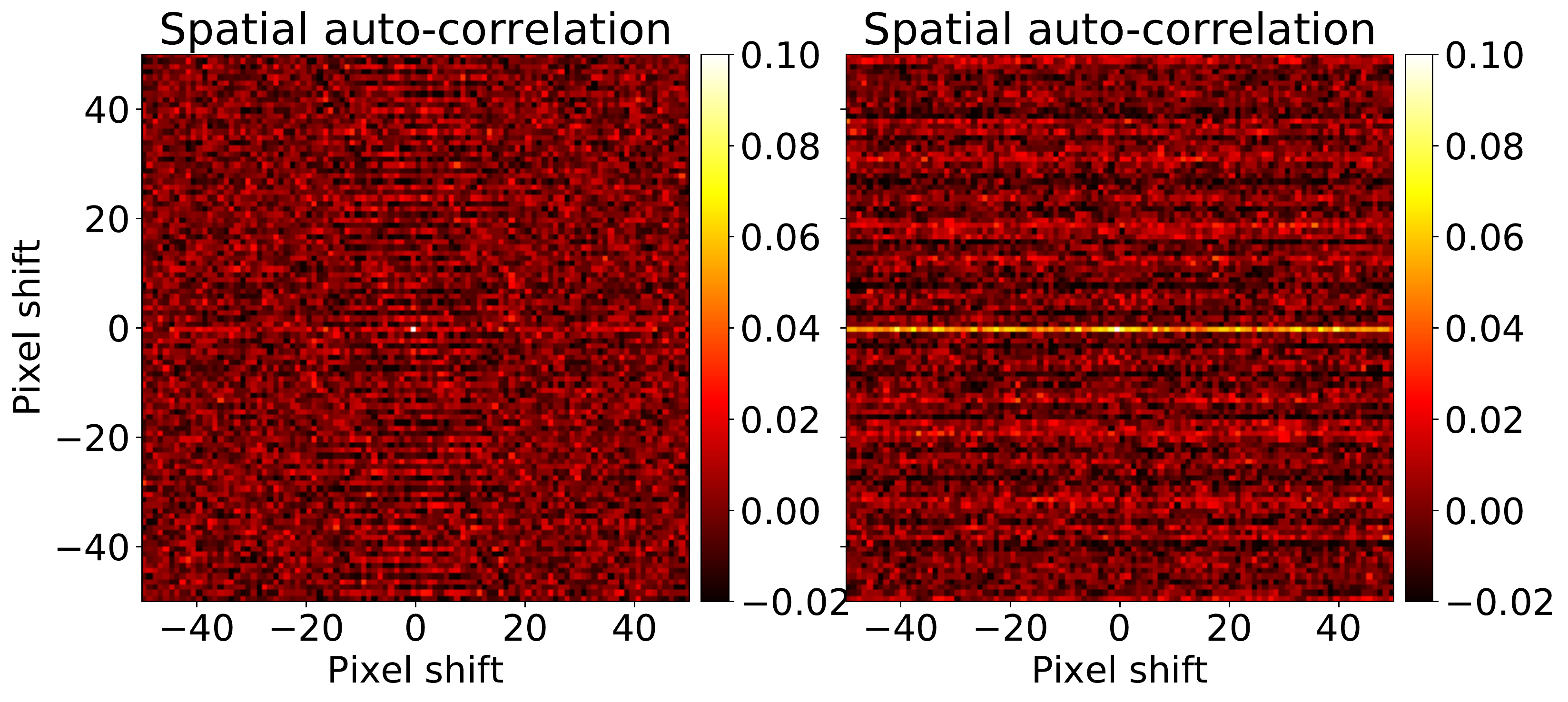}}
  }
  \caption{Mean spatial auto-correlation of an individual dark subtracted frames acquired on Andor Neo (left panel) and Andor Marana (right panel) in the same scale. Neo data shows a significant correlation along horizontal lines due to on-board overscan subtraction. Marana data shows no such effects of any significant amplitude, noise is practically spatially uncorrelated.
    Andor Neo data is courtesy of a Mini-MegaTORTORA project \citep{karpov_2019}.
    \label{fig_noise_corr}}
\end{figure}

\begin{figure}[t]
  \centerline{\resizebox*{1.0\columnwidth}{!}{\includegraphics[angle=0]{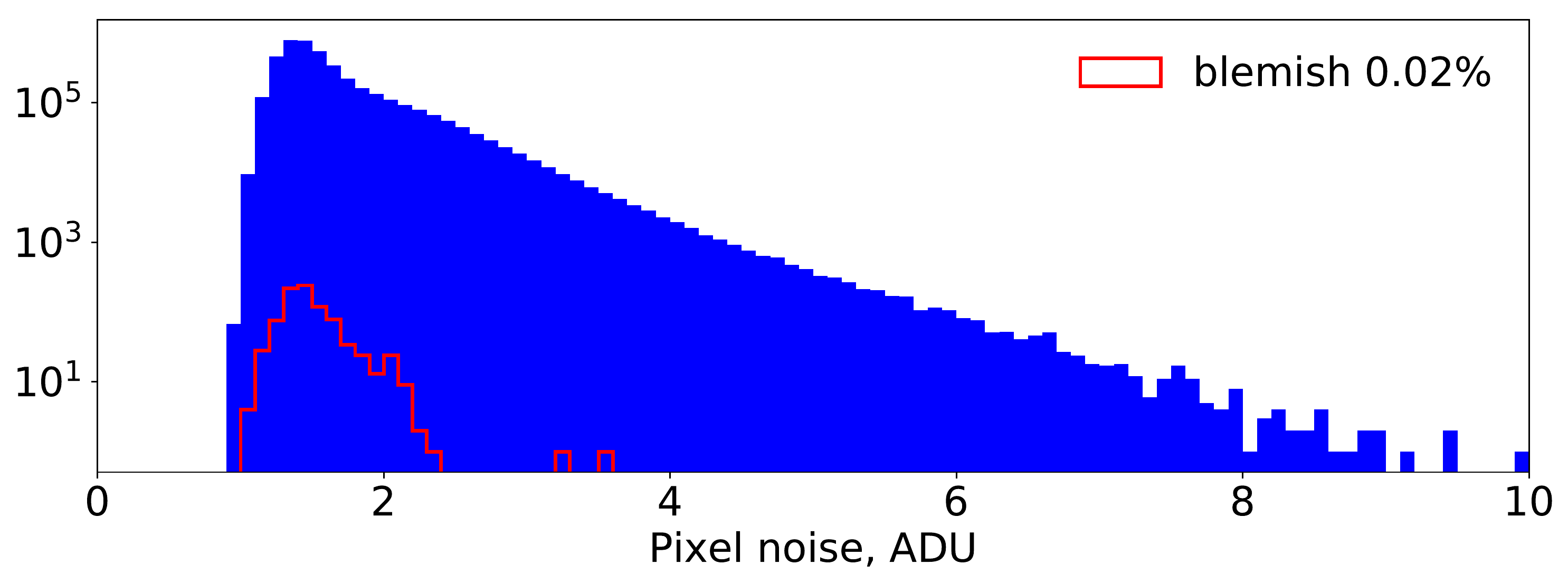}}}
  \caption{Histogram of a pixel noise RMS on a sequence of dark frames.
    \label{fig_noise_dark}}
\end{figure}

\begin{figure}[t]
  \centerline{\resizebox*{1.0\columnwidth}{!}{\includegraphics[angle=0]{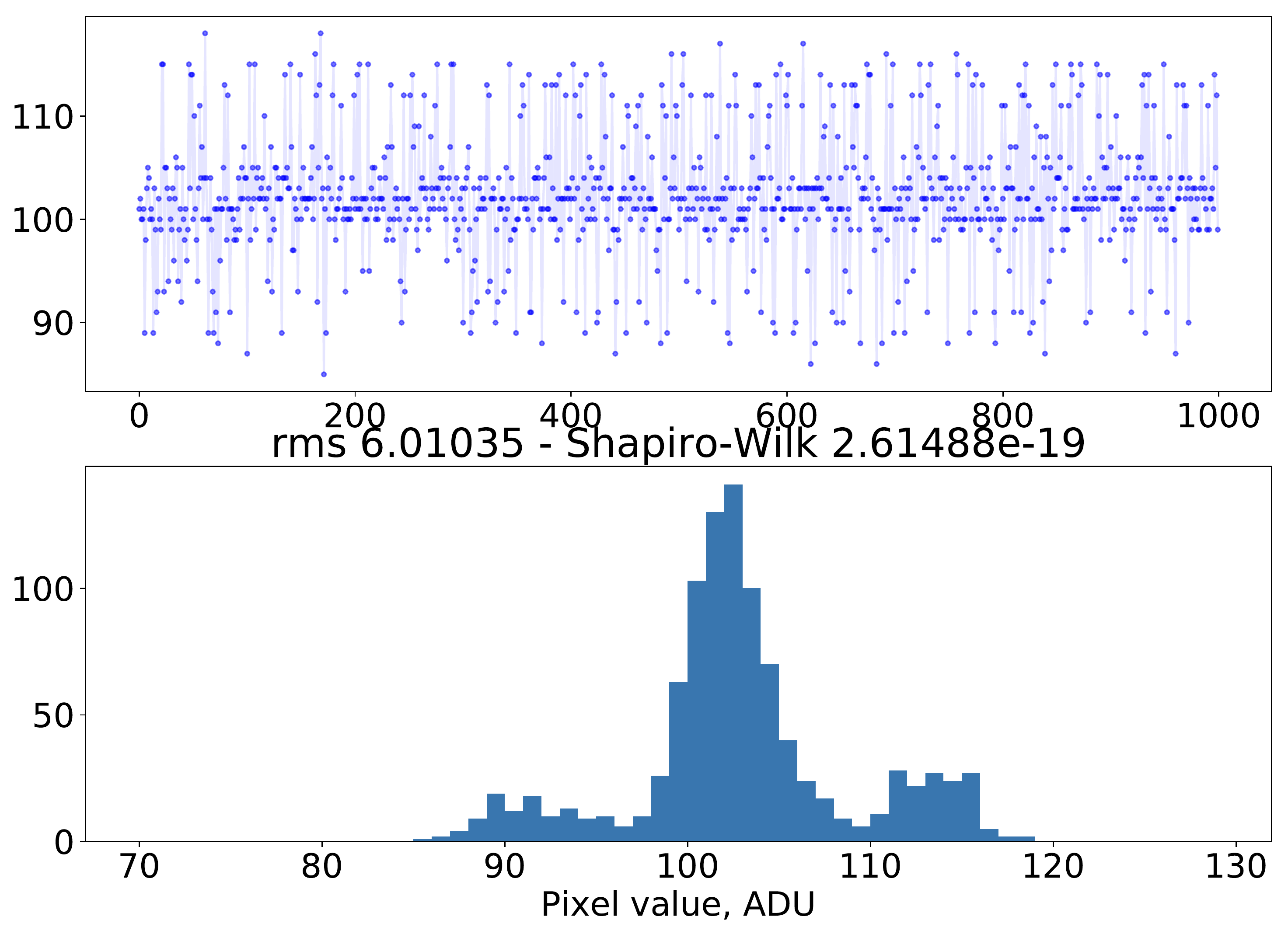}}}
  \caption{Example of a noisy pixel on a dark frame that displays distinctive Random Telegraph Signal (RTS) switching between several bias states. Upper panel -- temporal sequence of pixel values. Lower panel -- histogram of these values.
    \label{fig_noise_rts}}
\end{figure}

\begin{figure}[t]
  \centerline{\resizebox*{1.0\columnwidth}{!}{\includegraphics[angle=0]{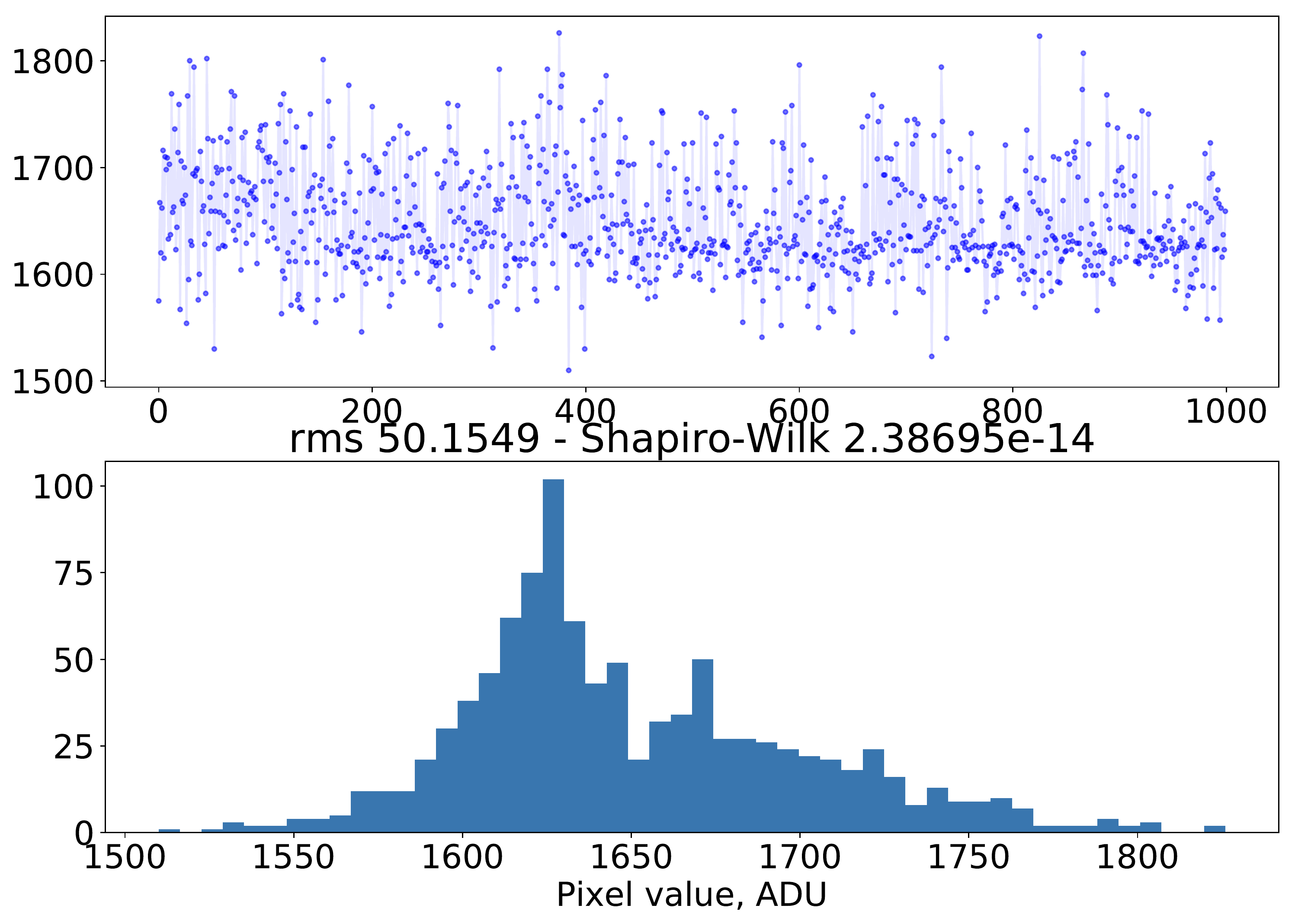}}}
  \caption{The same as Figure~\ref{fig_noise_rts}, but for a pixel at the amplifier transition intensity region. Excessive noise is caused by the ``jumping'' between the readings of high gain and low gain amplifiers.
    \label{fig_noise_transition}}
\end{figure}

In order to study the noise behaviour on a per-pixel basis, we studied the behaviour of every pixel pixel values over long consecutive sequences of frames, both dark and illuminated. The first point to note about the noise properties on Andor sCMOS cameras is an interpolative masking of a small subset of ``blemished'' pixels having either too high dark current or excess read-out noise. An on-board ``blemish correction'' \citep{neo_blemish} replaces the values of these pixels with an average of 8 surrounding ones on the fly, effectively leading to formation of a sub-set of pixels with noise level on average $\sqrt{8}$ times smaller than the rest. Direct search for such pixels may be performed by analyzing a series of illuminated images and locating the pixels with values always equal to arithmetic mean of its surroundings. The analysis of acquired data allowed us to identify 0.02\% of pixels as a blemish masked. The same  value for Andor Neo is typically around 0.66\%.

The noise on individual frames after subtraction of dark level is practically uncorrelated between adjacent pixels (see right panel of Figure~\ref{fig_noise_corr}). In contrast, the data from Andor Neo camera displayed a distinctive correlation along horizontal lines \citep{schildknecht_2013}, leading to a bit ``striped'' images. We suggest that the reason is on-board integer subtraction of sensor overscan values, which is absent in newer Marana chip due to differences in sensor architecture.

The histogram of a noise on dark frames (see Figure~\ref{fig_noise_dark}) show a significant power-law tail towards high RMS values, which consists of a pixels often displaying distinctive Random Telegraph Signal (RTS) features, effectively ``jumping'' between several metastable signal levels due to effect on electron traps inside the pixel circuits during either first or second readout during the correlated double sampling process \citep{noise_rts}. The example of such behaviour is shown in Figure~\ref{fig_noise_rts}.
Under illumination levels with Poisson noise exceeding the amplitude of these jumps (which is typically around 10 ADU for Marana sCMOS), these pixels behave normally and their histograms does not show any significant deviations from a Gaussian. Then, on approaching the amplifier transition region (see Figure~\ref{fig_ptc}), another effect appears, related to the switching between the readings of low gain and high gain amplifiers, effectively looking like a spontaneous jumping of the value -- see Figure~\ref{fig_noise_transition} for an example. Further increase of the intensity leads to only high gain readings being used, and pixel values are again stable.

\section{On-sky testing}\label{sec_sky}

\begin{figure}[t]
  \centerline{\resizebox*{1.0\columnwidth}{!}{\includegraphics[angle=0]{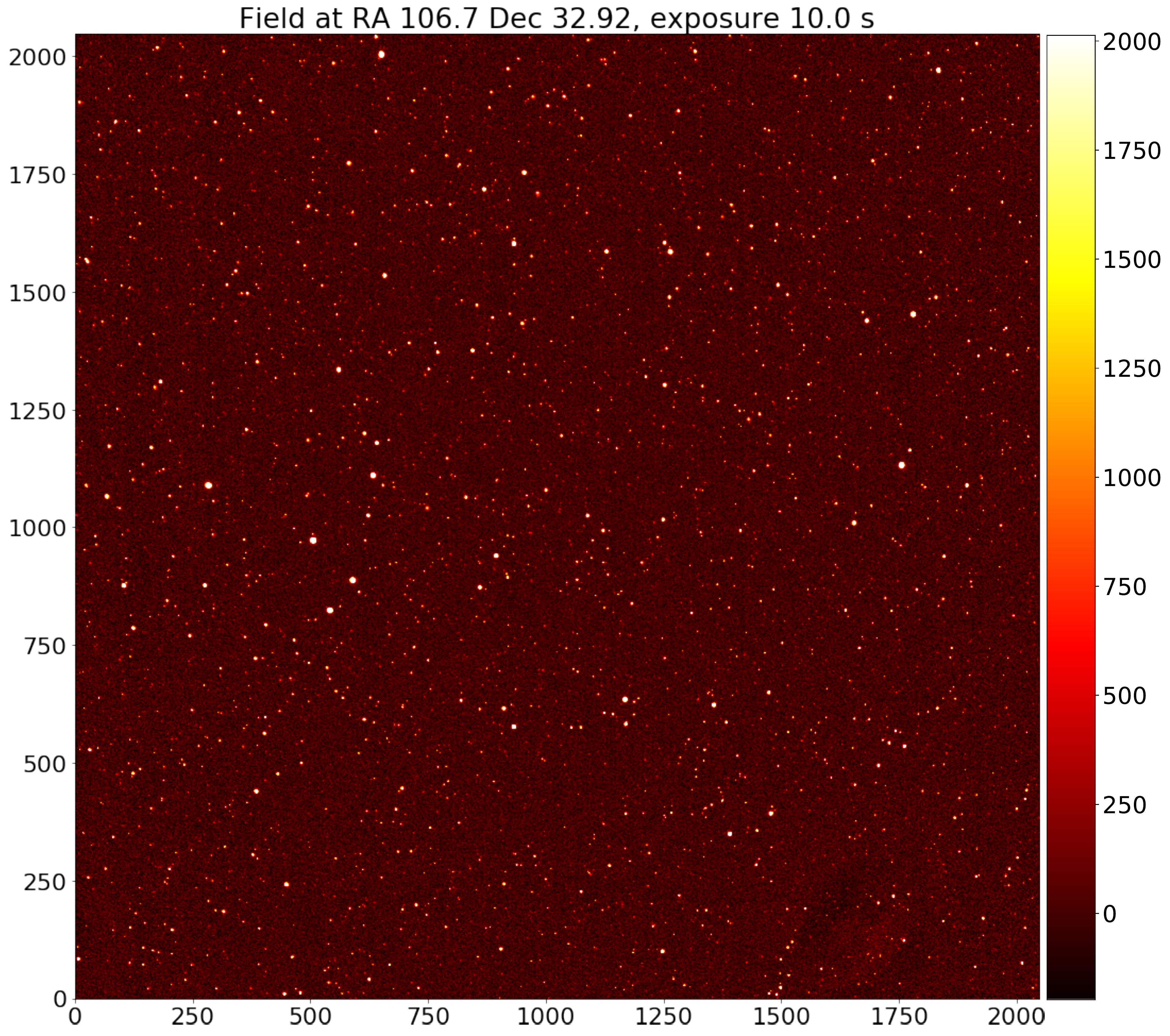}}}
  \caption{Single frame from on-sky testing of Andor Marana sCMOS, equipped with Nikkor 300 f/2.8 lens with no color filters. The frame is dark subtracted and flat-fielded using evening sky flat. The field of view is 4.26$^{\circ}$x4.26$^{\circ}$ with 7.5$''$/pixel scale. Median FWHM of the stars is 2.1 pixels, making the image nearly critically sampled. Note the absence of cosmetic defects typical for CCD frames -- hot and dark columns, bleedings from oversaturated stars, etc.
    \label{fig_sky}}
\end{figure}

\begin{figure}[t]
  \centerline{\resizebox*{1.0\columnwidth}{!}{\includegraphics[angle=0]{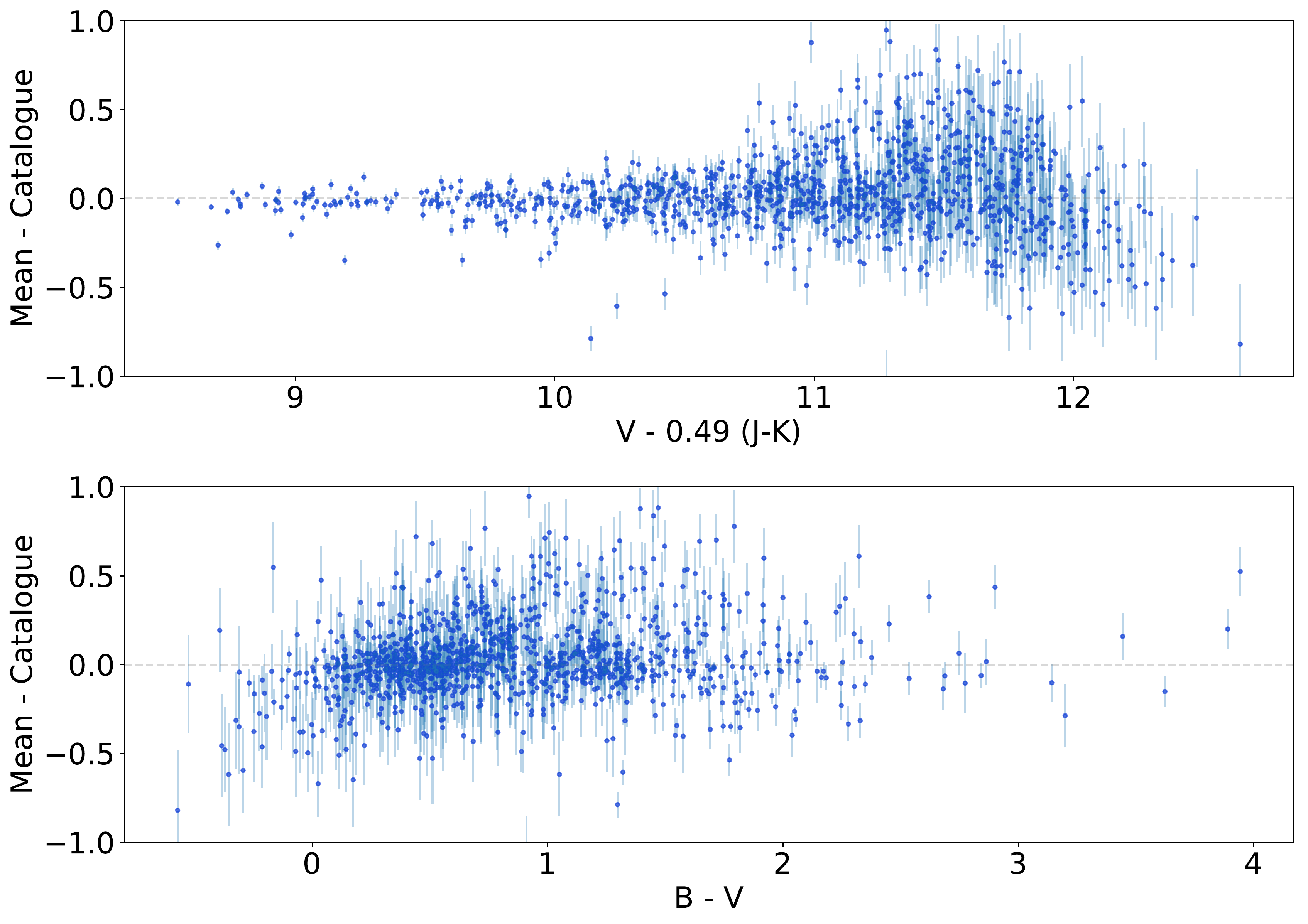}}}
  \caption{Difference of a mean magnitude measured along the sequence of sky images, and catalogue magnitude of corresponding star. Upper panel -- its dependence on magnitude, lower panel -- on catalogue star color. No significant systematic effects are seen. The error bars are catalogue magnitude errors; they are the primary source of large spread towards fainter stars.
    \label{fig_sky_diff}}
\end{figure}

\begin{figure}[t]
  \centerline{\resizebox*{1.0\columnwidth}{!}{\includegraphics[angle=0]{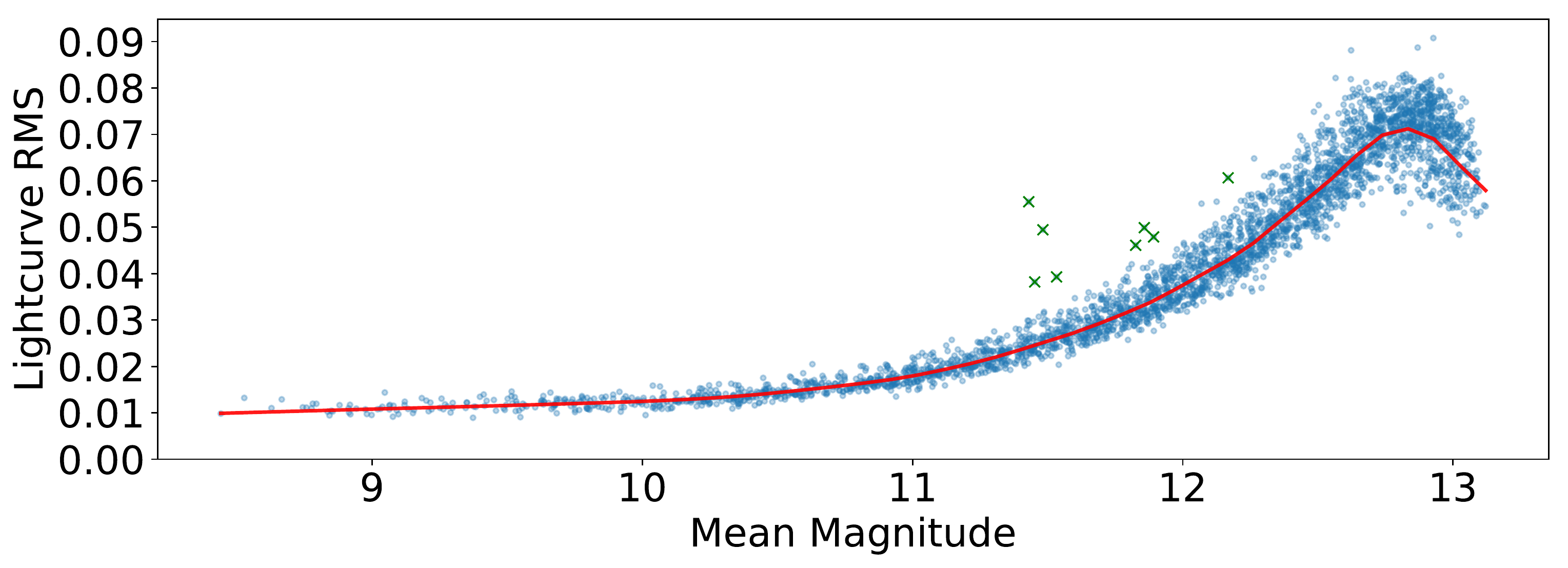}}}
  \caption{The scatter of photometric measurements of individual star along the sequence of sky images versus its mean value. A few outliers are caused by errorneous measurements of a blended stars.
    \label{fig_sky_rms}}
\end{figure}

On-sky testing of the camera consisted of a series of continuous observations of a fixed sky positions in order to assess the photometric performance and achievable stability of the data. Every frame (example is shown in Figure~\ref{fig_sky}) acquired in such regime was bias and dark subtracted and normalized to a flat field acquired by averaging evening sky images. Then every frame was astrometrically calibrated using {\sc Astrometry.Net} \citep{astrometry.net} code, and star detection and measurements were performed using routines available in {\sc SEP} \citep{SEP} Python package, based on original SExtractor code by \citet{sextractor}. On every frame, the zero point model was constructed by cross-matching the object list with the synthetic photometric catalogue of \citet{pickles} and fitting their instrumental magnitudes with a catalogue $V$ values as a base, $J-K$ as a color term, and a third order spatial polynomial to compensate imperfections of evening flats, as well as positional-dependent aperture correction due to changes of stellar PSF. By comparing the fits from different frames, the color equation for these unfiltered observations was found to be:
$$
\mbox{Instr} = V - 0.49\cdot(J-K)\ .
$$
After that, the color term was kept fixed for all frames, and the fitting was performed only for spatial polynomial part.  All photometric measurements of all stars on all frames were then positionally clustered and separated into light curves corresponding to individual objects. Then for every light curve the mean value and standard deviation were computed. The difference between these mean magnitudes and catalogue is shown in Figure~\ref{fig_sky_diff}, no systematics in either magnitude or color is visible there. The scatter versus magnitude plot for the lightcurves is shown in Figure~\ref{fig_sky_rms}, and demonstrates that the photometric precision of measurements with this sensor easily reaches 1\% in the setup we used. We did not detect any systematic effects dependent on sub-pixel position on the level greater 0.5\%, which is consistent with the chip being back-illuminated.

\section{Conclusion}\label{sec_conclusions}

The comparison of Andor Marana sCMOS to previous model, Andor Neo, according to manufacturer specifications is shown in Table~\ref{tab_marana}. Our results confirms that Marana dark current is indeed large, especially with on-board glow correction disabled.

The linearity of Marana is nearly perfect up to approximately half of saturation level, with just an occasional jump of typically less than 2\% at the amplifier transition region. The effective gain, however, drops  by nearly a two times there, also changing the spatial properties of flat fields above approx. 1500 ADU by revealing the structure of column level amplifiers. The behaviour is a bit better than the one in Neo.

The amount of blemished pixels in Marana is nearly 30 times smaller than in Neo, where they occupied up to 0.66\% of all pixel area and significantly affected the photometric performance. The noise at the dark level shows a random telegraph signal features in a small (0.01\%) with a jumps amplitude around 10 ADU. Moreover, the intensity region around amplifier transitions also shows a pixel value excess jumps related to switching between readings of low gain and high gain amplifiers. All other intensity regions show quite stable and normally distributed pixel readings over time.

Our on-sky tests of Marana display promising performance, with a photometric precision easily reaching 1\% on a sequence of consecutive frames even with critically sampled (FWHM$\approx$2 pixels) stellar profiles, and without any signs of sub-pixel position related systematics like the ones typically evident in Neo, which employs a microlens raster in front of a sensitive part of every pixel.

Moreover, the overall quality of the images are nice and free of typical CCD cosmetic problems -- hot and dark columns, bleedings from oversaturated stars, etc -- due to per-pixel read-out, which significantly increases the amount of scientifically usable pixels over the sensor area. Andor Marana also lacks the characteristic horizontal auto-correlation (striped images) evident in Neo frames due to on-board integer subtraction of overscans.

That all said, we may safely conclude that Andor Marana sCMOS is indeed a very promising camera for a sky survey applications, especially requiring high temporal resolution.

%\backmatter

\section*{Acknowledgments}

This work was supported by \fundingAgency{European Structural and Investment Fund} and the \fundingAgency{Czech Ministry of Education, Youth and Sports} (Project CoGraDS -- \fundingNumber{CZ.02.1.01/0.0/0.0/15 003/0000437}). Authors are grateful to Andor, Oxford Instruments Company for providing the camera used for testing, and to Kazan Federal University for providing test data from Andor Neo sCMOS cameras.

\subsection*{Author contributions}

S.K. created the software used for data acquisition, performed the acquisiton of test data, analysis of the data and discussion of its results. A.B. participated in the analysis of the data, preparation of hardware for the experiments and discussion of results. A.C. participated in preparation of hardware for the experiments and in discussion of results. M.P. negotiated the availability of the camera used for testing and overviewed the whole project.

\subsection*{Financial disclosure}

None reported.

\subsection*{Conflict of interest}

The authors declare no potential conflict of interests.

% \nocite{*}% Show all bib entries - both cited and uncited; comment this line to view only cited bib entries;
\bibliography{cmos}%

\section*{Author Biography}
% (if applicable)

\begin{biography}
  {\includegraphics[width=70pt,height=70pt,draft]{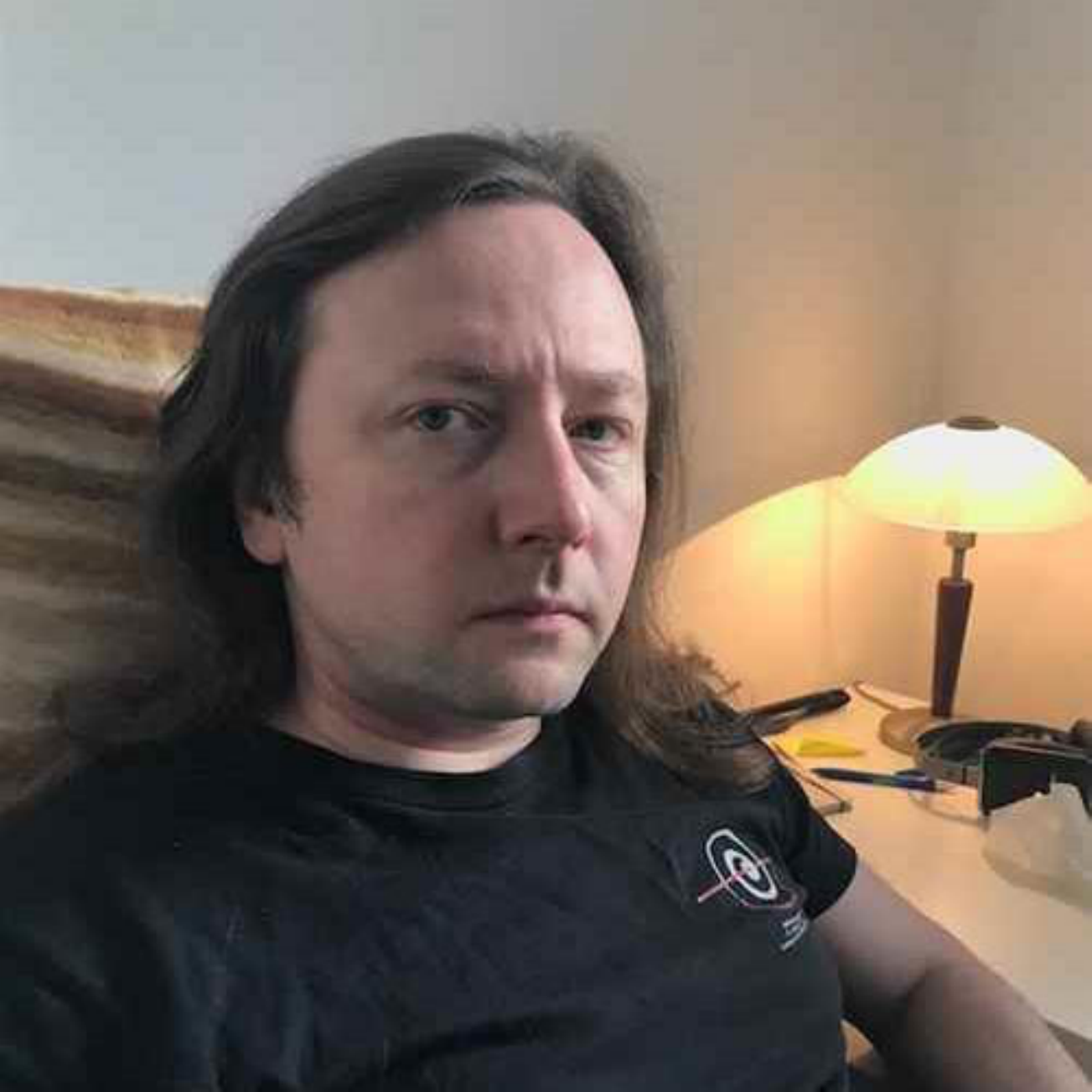}}
  % {\includegraphics[width=70pt,height=70pt]{fig_karpov}}
  {\textbf{Karpov, Sergey.}
Sergey Karpov finished Moscow State University in 2003, and got his PhD in astrophysics in Special Astrophysical Observatory, Russia, in 2007. His scientific interests include high temporal resolution astrophysics, time-domain sky surveys, astronomical data processing pipelines and transient detection algorighms. He is currently working on development of automated data analysis pipelines for several time domain sky surveys, as well as on various aspects of testing and characterization of optical detectors used in astronomy.}
\end{biography}

\end{document}